\documentclass[aps,pre,preprint,groupedaddress]{revtex4}
\usepackage{graphicx}

\begin{document}

\preprint{LA-UR-10-01844}

\title{Comparative Monte Carlo Efficiency by Monte Carlo Analysis}


\author{B. M. Rubenstein}
\affiliation{Chemistry Department, Columbia University, New York, NY 10027}

\author{J. E. Gubernatis}
\affiliation{Theoretical Division, Los Alamos National Laboratory, Los Alamos, NM 87545}

\author{J. D. Doll}
\affiliation{Chemistry Department, Brown University, Providence, RI 02912}


\date{\today}

\begin{abstract}
We propose a modified power method for computing the subdominant eigenvalue $\lambda_2$ of a matrix or continuous operator. While useful both deterministically and stochastically, we focus on defining simple Monte Carlo methods for its application. The methods presented use random walkers of mixed signs to represent the subdominant eigenfuction. Accordingly, the methods must cancel these signs properly in order to sample this eigenfunction faithfully. We present a simple procedure to solve this sign problem and then test our Monte Carlo methods by computing the $\lambda_2$ of various Markov chain transition matrices. As the $|\lambda_2|$ of this matrix controls the rate at which Monte Carlo sampling relaxes to a stationary condition, its computation also enabled us to compare efficiencies of several Monte Carlo algorithms as applied to two quite different types of problems. We first computed ${\lambda_2}$ for several one and two dimensional Ising models, which have a discrete phase space, and compared the relative efficiencies of the Metropolis and heat-bath algorithms as a function of temperature and applied magnetic field. Next, we computed $\lambda_2$ for a model of an interacting gas trapped by a harmonic potential, which has a mutidimensional continuous phase space, and studied the efficiency of the Metropolis algorithm as a function of temperature and the maximum allowable step size $\Delta$.  Based on the $\lambda_2$  criterion, we found for the Ising models that small lattices appear to give an adequate picture of comparative efficiency and that the heat-bath algorithm is more efficient than the Metropolis algorithm only at low temperatures where both algorithms are inefficient. For the harmonic trap problem, we found that the traditional rule-of-thumb of adjusting $\Delta$ so the Metropolis acceptance rate is around 50\% range is often sub-optimal. In general, as a function of temperature or $\Delta$, $\lambda_2$ for this model displayed trends defining optimal efficiency that the acceptance ratio does not. The cases studied also suggested that Monte Carlo simulations for a continuum model  are likely more efficient than those for a discretized version of the model.
\end{abstract}

%\keywords{}

\maketitle

\section{Introduction\label{}}
%\subsection{}
%\subsubsection{}

When designing a Monte Carlo simulation, the computational scientist often must decide which of several algorithmic options is the most efficient or how to optimize a particular algorithm.  Besides experience, few rules of guidance exist. 

For detailed balance algorithms, the class we assume, likely the most rigorous rule is based on the work of Peskun \cite{peskun73}.  He showed analytically that if $P^{1}$ and $P^{2}$ are two transition matrices  that satisfy detailed balance and asymptote to the same limiting distribution, then for jumps from state $j$ to state $i$,  algorithm 1 is more efficient than algorithm 2 if
\begin{equation}
P_{ij}^1 > P_{ij}^2,\text{ for all }i \ne j.
\label{eq:peskun}
\end{equation}
While he used this relation to establish the greater efficiency of the Metropolis-Hastings algorithm over several other generalized Metropolis algorithms, establishing this relation on a case by case basis is generally difficult for large phase spaces.

The theory of Markov chains says the magnitude of the transition matrix's subdominant eigenvalue $\lambda_2$ controls the rate at which the sampling relaxes to a stationary condition. Doll \emph{et al.\/} \cite{doll09} call this eigenvalue the asymptotic convergence parameter \cite{note1}. Because its magnitude must be less than one, its closeness to zero indicates a high degree of efficiency, and closeness to 1, poor efficiency.  Statisticians in particular \cite{liu01} have derived a number of upper and lower bonds for this eigenvalue.  Again, making use of this rigorous information is sometimes difficult.

Oral tradition says that efficient sampling occurs when the acceptance ratio is around 50\%. The acceptance ratio is the number of jumps to a different state divided by the total number of jumps attempted. In some sense tradition is consistent with Peskun. Because the transition matrix satisfies $\sum_i P_{ij}=1$, moving transition probability off of the diagonal generally increases the acceptance. However, the more relevant implication of Peskun's result is the need for jumps among areas distantly separated in phase space. Acceptance ratios are particularly misleading in cases where phase space separates into several relatively localized regions of large Boltzmann weight separated by an energy activation barrier larger than thermal fluctuations. In this circumstance, sampling is quasi-non-ergodically confined to one region, and within this region the acceptance ratio may be, or may have been adjusted to be, the canonical 50\%.

A direct approach to assessing the second eigenvalue is computing it numerically. The size of the transition matrix (or in many cases the transition operator) restricts a direct deterministic approach to small systems because of the amount of computer memory required. Recently, Doll \emph{et al.\/} \cite{doll09} used the deterministic approach on the Metropolis algorithm's transition matrix for single quadradtic, quartic, double well, and triple well oscillator potential energies. They discretized the one-dimensional phase space and prohibited very large displacements to create a finite matrix representation of the asymmetric transition matrix. They then used standard eigensystem software to compute the subdominant eigenvalue as a function of temperature and the Metropolis box size. The Metropolis box size is the maximum size of the proposed move from the current position in phase space.   

Doll \emph{et al.} found several interesting results. The results for the quadratic and quartic wells were similar: adjusting the box size so the acceptance rate is around 50\% resulted in the second eigenvalue being reasonably removed from unity.  For the double and triple wells, the competition between intra- and inter-well sampling made box size optimization more difficult. In some cases an acceptance rate around 50\%  corresponded to an unfavorable second eigenvalue.  Most interestingly, apparent activation energies in the behavior of the second eigenvalue existed and depended on the box size.  Further, as a function of box size, they found structure in the second eigenvalue reflecting the expected phase space structure. The acceptance ratio however featured little of the informative structure seen in the second eigenvalue. In short, the sampling dynamics, in particular the length scale dependence of the apparent activation energies, contained information about the underlying structure of the potential energy surface about the width of  the barrier region, for example.

Extending this type of deterministic analysis to higher dimensional phase spaces,  however, is rapidly checked by inadequate computer memory. Accordingly, we are proposing a shift to a new Monte Carlo method to make this extension possible.  We  will illustrate the usefulness of this new method by comparing the second eigenvalue for several Monte Carlo algorithms applied to the one and two-dimensional Ising models and for the Metropolis algorithm applied to a gas of $N$ interacting particles in a harmonic trap.

Computing the subdominant eigenpair $(\lambda_2,|\psi_2\rangle)$ is a significant shift beyond conventional Monte Carlo eigenanalysis. The Monte Carlo methods based on the power method have long computed just the dominant eigenpair $(\lambda_1,|\psi_1\rangle)$ of very large matrices. For the Monte Carlo transition matrices, the dominant eigenvalue is known and must be one. Further, the left and right hand dominant eigenvectors are also known. If $P$ is column stochastic, that is, $\sum_i P_{ij} =1$, the right eigenvector is the limiting distribution.  The left eigenvector has all its components being the same positive number. Our method based on a modified power method, uses this information about the dominant eigenpair to focus on obtaining the first subdominant pair. It starts with the recent very large matrix multiple eigenvalue Monte Carlo work of Booth and Gubernatis \cite{booth03,booth08,gubernatis08,booth09} and adds to it an algorithm just proposed by Yamamoto \cite{yamamoto09}. 

Our approach is significantly different from the one proposed by Bl\"ote and Nightingale \cite{nightingale96a, nightingale96b, nightingale00} who adapted the variational quantum Monte Carlo method \cite{hammond94} for computing a few excited states to the calculation of subdominant eigenvalues.  In their modificaton of this variational approach, they utilized knowledge of the dominant eigenpair, and with a multi-parameter trial wavefunction, accomplished high precision estimates of the dynamical exponent for a Metropolis simulation of the two-dimensional Ising model at the bulk critical point. Important in this method is the quality of the trial wavefunction.  Recently, Casey \emph{et al. }\cite{casey09} applied this method to a multi-variate Gaussian.

In Section 2, we will introduce our method and in Section 3 define our models and discuss the details of our simulations. Section 4 contains our results for the Ising and harmonic trap models. Because Ising model simulations have no Metropolis box size, we instead varied the algorithm, using the standard single spin-reversal Metropolis method, plus single- and multi-site heat bath algorithms. In the last section, Section 5, we summarize our results and discuss future work.

\section{Background}

Most commonly used Monte Carlo eigenvalue methods are based on the power method. The power method projects some starting state to the eigenpair of some matrix or operator $A$ associated with the eigenvalue of largest absolute value. For a Markov chain transition matrix $P$, this eigenvalue must always be real, positive, and unity. With an initial state $|\psi\rangle$, the power method iterates
\begin{eqnarray}
1. & &|\phi\rangle = A|\psi\rangle \nonumber \\
2. & &|\psi\rangle = |\phi\rangle/\|\phi\|
\end{eqnarray}
until some convergence criterion is met. Upon convergence, the eigenstate of the dominant eigenpair is $|\phi\rangle$ and the eigenvalue is $\| \phi \|$. Any norm may be used. In this paper, we will use the infinity norm,  $\| \phi \| =\max_{i}  | \phi_i | $,  where $\phi_i$ are the components of $|\phi\rangle$ in some basis.

To find two eigenpairs, we need two starting states $|\phi'\rangle$ and $|\phi''\rangle$. If we were to apply the power method to them independently, each would independently converge to the same dominant eigenpair. To couple the iteration, we modify the first step of the power method to be
\begin{eqnarray}
|\psi'\rangle &=&| \phi'\rangle + \eta_1 |\phi''\rangle \nonumber \\
|\psi''\rangle &=&| \phi'\rangle + \eta_2 |\phi''\rangle,  
\label{eq:power_method}
\end{eqnarray}
with the intent of converging $|\psi'\rangle$ to the dominate state $|\psi_1\rangle$ and $|\psi''\rangle$ to the subdominant state $|\psi_2\rangle$.

To fix values for the $\eta$'s, we start with the matrix-vector form of the defining equation for an eigenpair
\begin{equation}
\lambda\psi_i=\sum_j A_{ij} \psi_j
\end{equation}
and note that for any $\psi_i \ne 0$ an exact estimator for the eigenvalue is
\begin{equation}
\lambda =\frac{\sum_j A_{ij} \psi_j}{\psi_i}
\label{eq:estimator}
\end{equation}
and related exact estimators exist for sums of these components grouped in any number of overlapping or non-overlapping ways
\begin{equation}
\lambda =\frac{\sum_{i \in R_1}\sum_j A_{ij} \psi_j}{\sum_{i \in R_1}\psi_i}
=\frac{\sum_{i \in R_2}\sum_j A_{ij} \psi_j}{\sum_{i \in R_2}\psi_i}
=\dots=\frac{\sum_{i \in R_N}\sum_j A_{ij} \psi_j}{\sum_{i \in R_N}\psi_i}.
\label{eq:region_estimator}
\end{equation}
This observation puts a constraint on the allowed values of $\eta$ for which the sum $\psi_i=\phi'_i +\eta \phi''_i$ is an eigenvector. If we substitute, $\psi_i$ into (\ref{eq:region_estimator}) and cross-muliply the equalities involving $R_1$ and $R_2$, a quadratic equation for $\eta$ results. One root of this equation, $\eta_1$, makes the eigenvalue estimate $\lambda_1$ associated with $\phi'_i +\eta \phi''_i$ larger than the other. As shown in \cite{gubernatis08,booth09}, these choices  converge $|\psi'\rangle$ and $|\psi''\rangle$ to $|\psi_1\rangle$ and $|\psi_2\rangle$. It is straight-forward to generalize this method to compute more that two eigenpairs.

Following Yamamoto \cite{yamamoto09}, we modified this method to converge to the subdominant eigenpair using knowledge of the first. We define
\begin{equation}
\begin{array}{*{20}l}
   {P_1  = \sum\limits_{i \in R_1 } {\phi _i } } & \quad {P_2  = \sum\limits_{i \in R_2 } {\phi _i } }  \nonumber \\
   {P'_1  = \sum\limits_{i \in R_1 } {\sum\limits_j {A_{ij} \phi _j } } } & \quad {P'_2  = \sum\limits_{i \in R_2 } {\sum\limits_j {A_{ij} \phi _j } } }.  \\
\end{array}
\end{equation}
After a sufficient number of iterations
\begin{eqnarray}
|\psi'\rangle   &\sim&  a_1|\phi_1\rangle +a_2|\phi_2\rangle                                       \sim  a_1|\psi_1\rangle +a_2|\psi_2\rangle\\
|\psi'' \rangle &\sim&  a_1\lambda_1|\phi_1\rangle +a_2\lambda_2|\phi_2\rangle \sim  a_1\lambda_1|\psi_1\rangle +a_2\lambda_2|\psi_2\rangle.
\end{eqnarray}
Accordingly, we can write
\begin{equation}
\begin{array}{*{20}c}
   {P_1  \sim a_1 \alpha _1  + a_2 \beta _1 } &\quad  {P'_1  \sim a_1 \alpha _1 \lambda _1  + a_2 \beta _1 \lambda _2 }  \\
   {P_2  \sim a_1 \alpha _2  + a_2 \beta _2 } &\quad  {P_2 ^\prime   \sim a_1 \alpha _2 \lambda _1  + a_2 \beta _2 \lambda _2 }  \\
\end{array}
\label{eq:new_estimates}
\end{equation}
where 
\begin{equation}
\begin{array}{*{20}c}
   {\alpha _{1,2}  = \sum\limits_{i \in R_{1,2} } {\phi _{1,i} } } &\quad  {\beta _{1,2}  = \sum\limits_{i \in R_{1,2} } {\phi _{2,i} } }.  \\
\end{array}
\end{equation}
We can now solve (\ref{eq:new_estimates}) for
\[
\lambda _2  = \frac{{\alpha _1 P'_2  - \alpha _2 P'_1 }}{{\alpha _1 P_2  - \alpha _2 P_1 }}
\]
and 
\[
a_1  = \frac{{\lambda _2 P_1  - P'}}{{\alpha _1 \left( {\lambda _2  - \lambda _1 } \right)}}.
\]
The $P$'s and $\alpha$'s are easily computed sums. In computing $\alpha_1$ we use our exact knowledge of $|\psi_1\rangle$, and in computing $P'$, we use the exact value of $\lambda_1$ which is unity. The new power method iterates
\begin{eqnarray}
1.&  &|\psi\rangle = A|\phi\rangle \nonumber \\
2.&  &\text{Calculate } a_1 \text{ and } \lambda_2 \nonumber \\
3.&  &|\psi\rangle \leftarrow |\psi\rangle -\eta a_1 |\psi_1\rangle \nonumber \\
4.&  &|\phi\rangle = |\psi\rangle/ \|\psi\|
\label{eq:modified_power_method}
\end{eqnarray}
%\begin{equation}
%\text{
%\begin{enumerate}
%\item $|\psi\rangle = A|\phi\rangle$
%\item Calculate $a_1$ and $\lambda_2$
%\item $|\psi\rangle \leftarrow |\psi\rangle -\eta a_1 |\psi_1\rangle$
%\item $|\phi\rangle =(\text{Sign } \psi_1) |\psi\rangle/ \|\psi\|$
%\end{enumerate}
%}
%\label{eq:modified_power_method}
%\end{equation}
Yamamoto shows that his procedure converges to $\lambda_2$ and $|\psi_2\rangle$ provided
\begin{equation}
(\lambda_1-\lambda_2)/\lambda_1 <  \eta  < (\lambda_1+\lambda_2)/\lambda_1 .
\end{equation}
In what follows, we chose $\eta$ to be close to the lower bound.

Both deterministic and Monte Carlo use of this modified power method are possible. In the following section, we will give more details of our Monte Carlo implementation. In this implementation, we use the \textit{left} dominant eigenvector of $A$, that is, the one with uniform positive components. This choice trivializes the computation of several sums.

\section{Models and Methods}

%We considered two models. The Ising model provided a discrete transition matrix with a finite number of transition elements, while a gas in a harmonic trap model provided a continuous transition kernel with an infinite number of transition elements. The core of our Monte Carlo eigenpair method is the same in both cases.

\subsection{Ising Models}

We considered both the $L$ site one-dimensional model
\begin{equation}
E=-\sum_{i=1}^L(Js_i s_{i+1} +Hs_i)
\label{eq:ising_1d}
\end{equation}
and the $L \times L$ site two-dimensional model
\begin{equation}
E=-\sum_{i=1}^L  \sum_{j=1}^L (s_{i,j}s_{i+1,j}+s_{i,j}s_{i,j+1}+Hs_{i,j})
\label{eq:ising_2d}
\end{equation}
in an external field $H$. The Ising variables $s_i=\pm 1$. We assumed periodic boundary conditions; that is, $s_{i+L}=s_i$ in one dimension, and $s_{i+L,j}=s_{i,j+L}=s_{i,j}=s_{i+L,j+L}$ in two dimensions. 

For these models, we computed the second eigenvalue for the transition matrices of multiple Monte Carlo algorithms: the single-site, spin-reversal Metropolis algorithm, a single-site heat bath algorithm, and the two and three-site  heat bath algorithms. The algorithms are defined by the transition probability matrix $P_{S'S}$, which in turn defines the probability of jumping from state $|S\rangle$ to $|S'\rangle$. With $N$ equal to $L$ or $L^2$, our state is
\begin{equation}
|S\rangle=|s_1,s_2,\dots,s_N\rangle.
\label{eq:spin_configuration}
\end{equation}
The jump to state $|S'\rangle$ produced by the Metropolis algorithm has a proposed flip of the Ising spin at one site accepted or rejected according to
\begin{equation}
P_{S'S}=\min\left[1,\exp(-E(S')/kT)/\exp(-E(S)/kT)\right].
\end{equation}

The single-site heat bath algorithm transitions the state to itself or to one with the spin reversed at one site. If $E(\bar{S})$ is the energy of the state with the single spin reversed and all remaining spins fixed  and $Z=\exp(-E(S)/kT)+\exp(-E(\bar{S})/kT)$, then the non-zero elements of $P$ are
\begin{eqnarray}
P_{SS}  &=&  \exp ( - E(S)/kT)/Z \nonumber \\
P_{\bar{S}S} & =& \exp ( - E(\bar{S})/kT)/Z.
\end{eqnarray}
A multiple-site heat-bath algorithm is a natural extension of the above. The single-site heat-bath algorithm samples one of the  two  spin states from the conditional Boltzmann distribution of a single spin with all the other spin values fixed. A multiple-site heat-bath algorithm samples the state of several neighboring spins with the rest fixed. For an $n$-site algorithm one of $2^n$ states is selected.

For each of these algorithms, we computed the second eigenvalue deterministically using standard eigensystem software for small lattices, deterministically using (\ref{eq:modified_power_method}) for slightly larger lattices, and stochastically using (\ref{eq:modified_power_method}) for still larger lattices. We will now give the details of the Monte Carlo approach.

The Monte Carlo method implements the modified power method by using a collection of $M$ random walkers, each specified by a weight $w_S$ and a spin configuration state $|S\rangle$. The number of these states is $2^N$.  We represented a spin configuration by an integer $S$ in the range 0 to $2^{N-1}$ where each bit of this integer corresponds to a lattice site and a plus Ising spin maps to a set bit and a minus spin maps to an unset one. The weight represents the component of the subdominant eigenstate in the spin-configuration basis (\ref{eq:spin_configuration}). Monte Carlo becomes necessary when this number is too large for a deterministic calculation. The Monte Carlo method is most powerful when $M\ll 2^N$.  

The algorithm estimates the eigenvalue by using the walkers in the exact estimators (\ref{eq:estimator}) and (\ref{eq:region_estimator}) not by using them in a variational estimator \cite{nightingale96b,nightingale00}. This use needs two regions. In general, the choice of regions is not critical \cite{note2}, as long as they are populated by a sufficient number of walkers so that the walkers are representative of the eigenstate in that region. Regions therefore do not necessarily have to be exclusive and can overlap. For the zero-field Ising model,  our regions were all states with an up-spin majority and all states with a down-spin majority. We note this means states with equal numbers of up and down spins do not contribute to the estimation. This choice accommodated the fact that as the lattice size increases most of the walkers at low temperature are either in the all spins-up or all spins-down state. In the non-zero-field simulations, one region was the one state with all spins aligned with the magnetic field and several adjacent nearly all-aligned states (the number of adjacent states that needed to be included with the all-aligned state scaled with system size); the other region was all other states. 

In contrast to the dominant eigenstate, which must have only non-negative components, the subdominant eigenstate must have some negative components. While not essential, it is helpful if the initial walker weights mix plus and minus signs. 
%As exhibited by the exact solution of the zero-field models, or easily observed from the deterministic solution for small lattices, the signs of the components of a states and its spin-reversed partner are opposite. It is also helpful to build this fact into the initialization. Because the iteration subtracts the dominant state from the converging subdominant state, it can reverse the weight of the walkers especially in the earlier steps.  To eliminate this, it is useful to fix the signs of the walker weights relative to the fixed sign of a particular component subdominant  eigenstate component. This amount to replacing step 4 of (\ref{eq:modified_power_method}) by
%\begin{equation}
%4. |\phi\rangle = \text{Sign}(\psi_1)|\psi\rangle/ \|\psi\|
%\end{equation}
%where we fixed the sign relative to the first component of $|\psi\rangle$. 
While use of starting states whose variational energy is a significantly better approximate to the answer is possible, their use was unnecessary for the present study. 

For each walker in state $|S \rangle$, we need to sample $P_{S'S}$ to produce a walker in the state $|S' \rangle$. To do this we used two lists, one for the current walkers and one for the new ones produced by a Monte Carlo method that samples a state $|S' \rangle$ from the cumulative probability function of $P_{S'S}$, that is, from $\sum_{S''=0}^{S'} P_{S''S}$. With this procedure we produce one new walker for every old one.

%\emph{Note: I did not use the cumulative distribution because it seemed not to work. I believe this is because the eigenvector was not faithfully distributed to the walkers as the CDF changes the ratio of the number of different components distributed to the walkers. By this I mean that the eigenvector may have 1 small component and 1 very large component. Both components must be represented in an equal ratio via the walker weights. Using the CDF purposefully destroys this ratio and this creates a problem. Perhaps I could do this better, but I believe the issue is fundamental}

The third step in the algorithm, which updates the discrete components $\psi_S$ of $|\psi\rangle$ after adding or removing a contribution from the corresponding component of the known dominant eigenstate, can mix oppositely signed walkers contributing to the same state $|S \rangle$. It is essential for a correct solution to have oppositely-signed walkers cancel properly. Instead of using the weight cancellation methods developed in \cite{booth03,booth09,booth09b}, we tried a simpler method: After the generation of the new list of walkers is complete,  we scan it, identify all walkers in the same state, that is, identified by the same positive integer, and replace them with one walker whose weight is the sum of contributing walkers' weights. This list compression procedure worked well.

As the iteration progresses, the repeated matrix-vector multiplication creates some very large-weighted walkers and some very small-weighted ones. It is inefficient to process the small-weighted ones. Accordingly, after we compressed our list, we eliminated the small-weighted ones by a weight cut-off procedure: We scan the list, and if a $|w_S|$ fell below $\epsilon$, then we would draw a random number $\xi$. If $\xi$ was larger than $|w_S|/\epsilon$, we would keep this walker but would increase its weight to $w_S/\epsilon$. Otherwise, we would remove it from the list. We used $\epsilon=10^{-3}$. The weight cut-off procedure reduces the number of walkers. If reduced too much, say by a half, we replenished the walker population size by replacing the largest-weighted walkers with $m={\rm{integer }}(w_S+\xi)$  walkers with weight $w_S/m$ until the list size was approximately restored to its original size. 

\subsection{Harmonic Trap}

For models in the continuum, we considered a gas of $N$ interacting classical particles in a harmonic trap. We intend this to be a simple model of a ``cold atom'' system. The potential energy is 
\begin{equation}
V_{Trap}(X) ={1 \over 2}  \sum_{i=1}^N  \left[  K  x_i^2 +\sum_{j(\ne i)=1}^Nv(x_i-x_j) \right] 
\label{eq:harmonic_trap}
\end{equation}
Here, $X$ denotes a position in phase space, that is, $X=(x_1,x_2,\dots,x_N)$. The $x_i$ are the particle displacements from the trap center.  We chose $v(x_i-x_j)=C(|x_i-x_j|-d)^2$, a rather artificial interaction but one staging a competition between the tendency of particles to roam freely and independently within the externally-generated harmonic trapping potential and the tendency to position themselves within a distance of $d$ of each other to accommodate their mutual interactions. We always chose $d=1$.

For this model, we computed the second eigenvalue for the Metropolis algorithm. Here,  the transition probability for jumping from position $X$ to $X'$ is
\begin{equation}
P(X',X)=T(X',X)\min[1,\exp(-V(X')/kT)/\exp(-V(X)/kT)].
\end{equation}
The Metropolis algorithm proposes a phase space position change one particle at a time, say changing $x_i$ to $x'_i$. The proposal samples an $x'_i$ from $T(X',X)=\prod_{i=1}^N t(x'_i,x_i)$ where $t(x'_i,x_i)=0$ unless $|x'_i|=|x_i +\Delta| \leq \Gamma$, where $\Gamma$ is a cutoff appropriate to the potential. The random number $\xi$ is chosen uniformly in the interval $(-\Delta,\Delta)$. The parameter $\Delta$ is called the Metropolis box size. The proposed change is accepted if $\exp(-V(X')/kT)/\exp(-V(X)/kT)$ is greater than another random number $\xi$ chosen uniformly in $[0,1]$. The parameters $\Gamma$ and $\Delta$ may be varied widely. Their ratio is what principally affects the second eigenvalues obtained. 

Our Monte Carlo strategy was analogous to the one we used for the Ising model, but some details were adjusted to move from discrete to continuous states. In short, we mapped the continuum problem onto a discrete one. We represented each walker by a  weight $w_X$ and phase space position $X$, divided phase space into cells, and formed cell lists grouping walkers into the cells. We then defined components for the various $|\psi\rangle$ and $|\phi\rangle$ states by mapping a combination of cell numbers and particle numbers onto a positive integer. We  thus defined discrete states, but ones typically containing many walkers. Finally, the weights of the walkers in each state were replaced by their average weight.   Except where otherwise noted, the cell width in each phase space dimension was 0.05. We observed that we could reduce the needed number of walkers by varying the widths. The number of walkers expected to occupy cells at the extreme ends of the trap, for example, is much smaller than the number expected to occupy cells at the bottom of the trap. Therefore, enlarging the cell widths above 0.05 units at the far ends of the trap enabled a reduction in the number of walkers. The optimal size of the cells at the far ends of the trap varied with the temperature, but could be optimized by increasing the size of these cells to the maximum size before convergence could no longer be achieved. We found these sizes ranged from 0.05 to 0.5 times the core cell width. The infinity norm of the eigenvector needed in step 5 of the algorithm was now taken to be the largest absolute value of the cell weights. 

We also need to define regions to calculate the parameters needed in the updating step 3. Our definitions were simple: The first region included all walkers with all $x_i < 0$,  while the second region included all walkers with all $x_i>0$. Other choices of regions are possible, but were not thoroughly explored.  Our weight control and population resizing procedures were the same as those used in the Ising simulations. Assigning each walker in the same cell their average weight was our only weight cancellation procedure.

\begin{figure}[t]
\begin{center}
\includegraphics[width=2.5in]{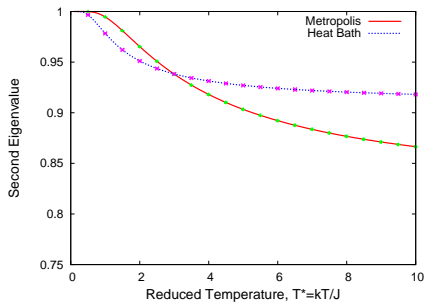}
\caption{Comparison of the deterministic and Monte Carlo calculations of the second eigenvalues of the Metropolis and single-site heat bath transition matrices for a zero filed one-dimensional Ising model on a 10 site lattice. }
\label{fig1}
\end{center}
\end{figure}

\begin{figure}[t]
\begin{center}
\includegraphics[width=3in]{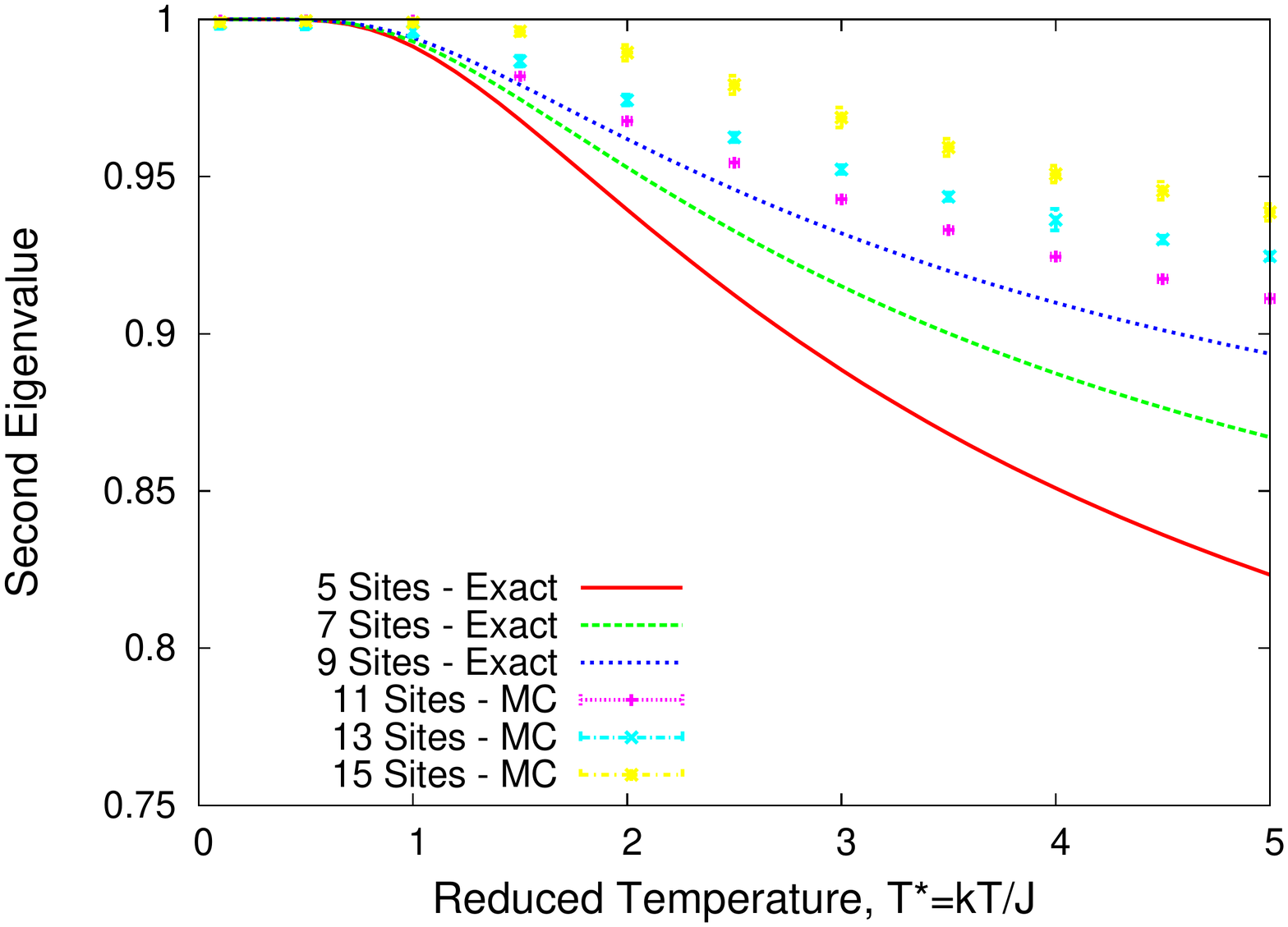}
\includegraphics[width=3in]{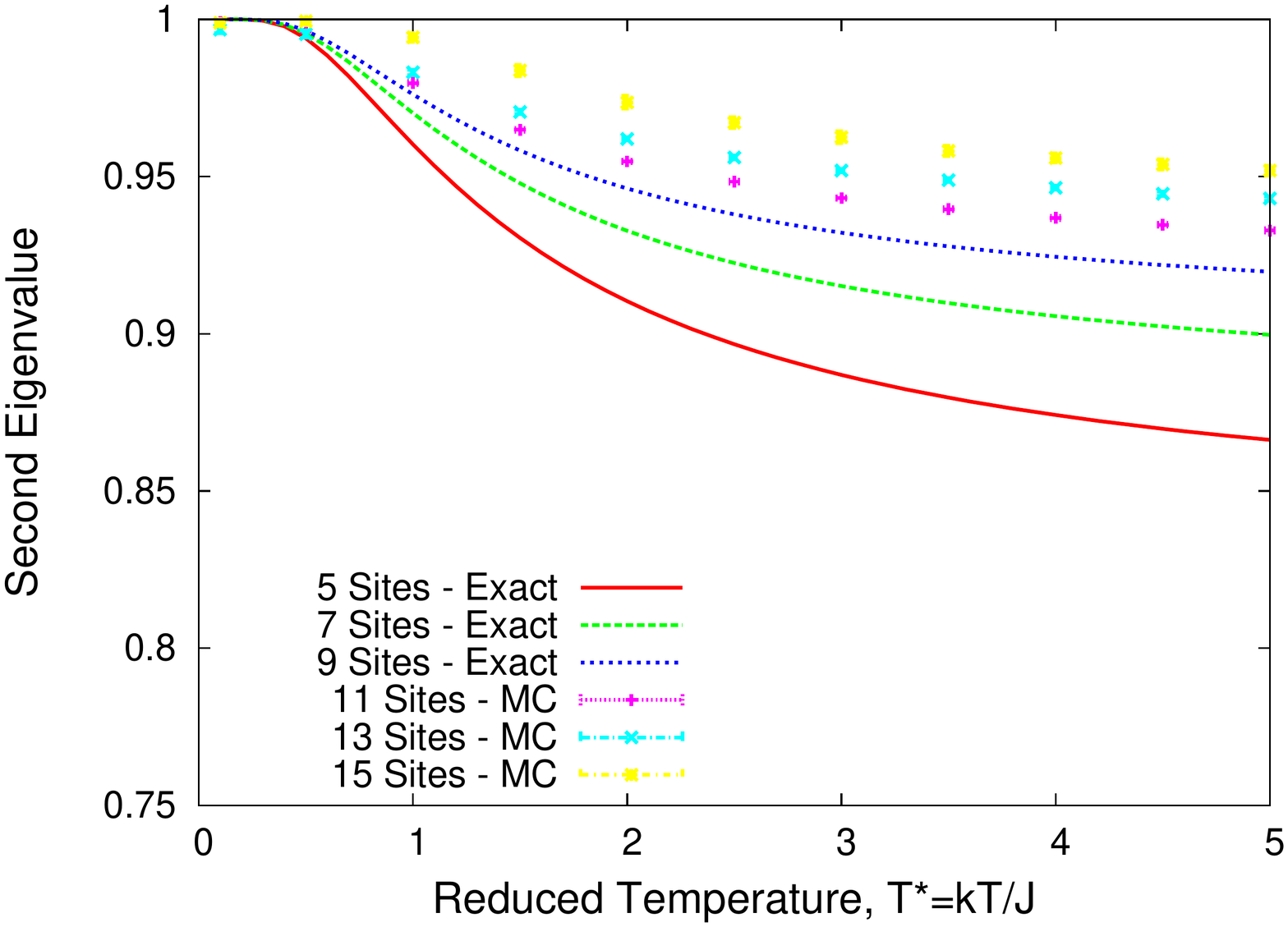}
\caption{Second eigenvalues for the Metropolis (left) and single-site heat bath (right) transition matrices for one-dimensional Ising lattices in zero magnetic field computed deterministically and by the proposed Monte Carlo method.}
\label{fig2}
\end{center}
\end{figure}

\section{Results}

\subsection{Ising Models}

Figure \ref{fig1} is representative of the excellent fidelity of our Monte Carlo eigenvalue predictions. Here,  as a function of the reduced temperature, we compare the Metropolis's and single-site heat bath's algorithms subdominant eigenvalue for a zero-field, 10-site lattice, computed by both deterministic (using standard eigenvalue software) and Monte Carlo approaches. The results for other lattice sizes and for two-site and three-site heat-bath algorithms display similar accuracy. All the deterministic calculations we present were done by using standard software. Computing the eigenvalue by (\ref{eq:modified_power_method}) gave the same result.

\begin{figure}[b]
\begin{center}
\includegraphics[width=3in]{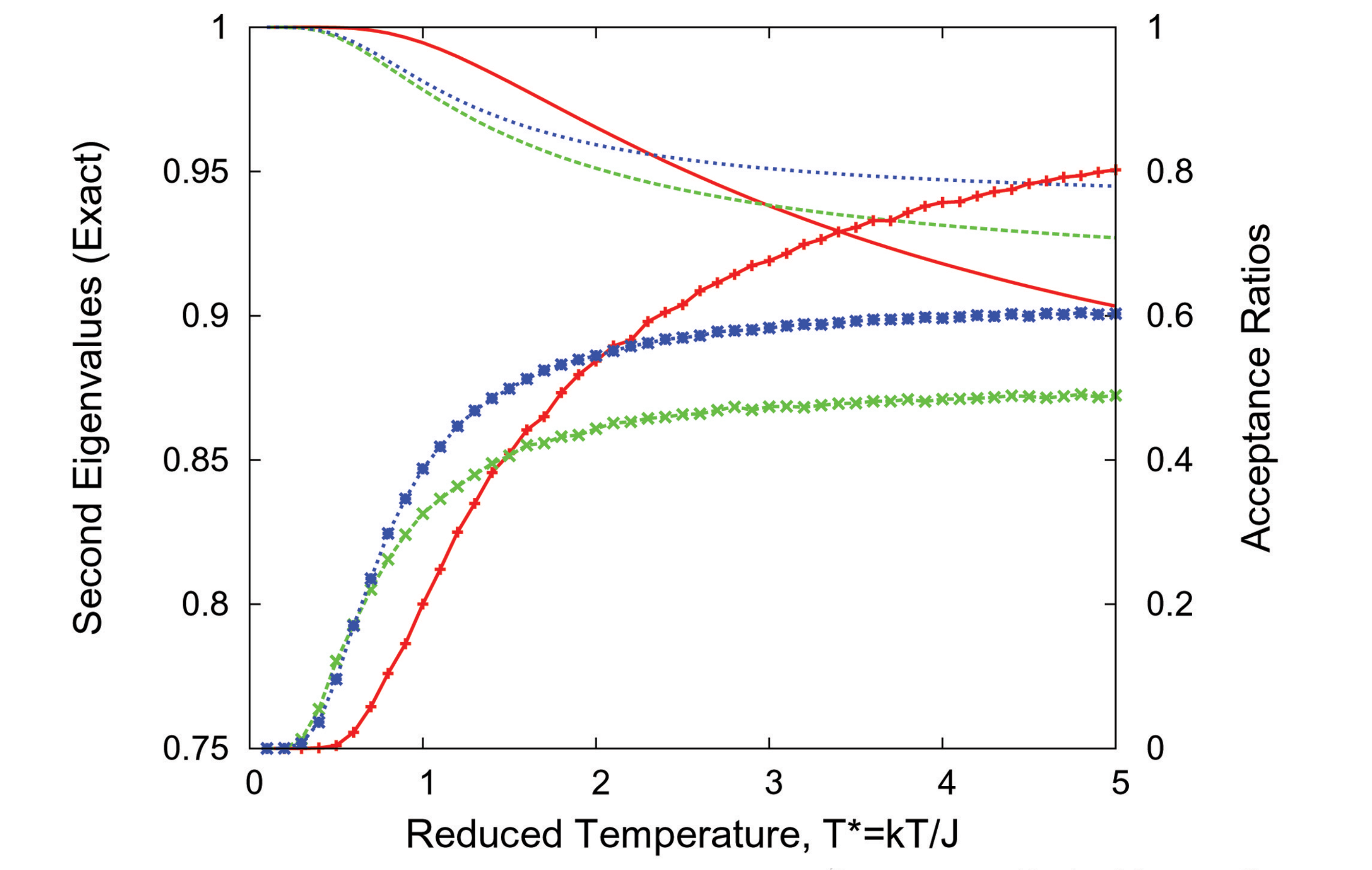}
\caption{Comparisons of the deterministic calculations of second eigenvalues and of the acceptance ratios for the Metropolis, single-site heat bath, and two-site heat bath algorithms for a one-dimensional Ising 10-site lattice in zero magnetic field. Red is for the Metropolis algorithm; blue, for the single-site heat bath; and green, the two-site heat bath.}
\label{fig3}
\end{center}
\end{figure}

\begin{figure}[t]
\begin{center}
\includegraphics[width=3in]{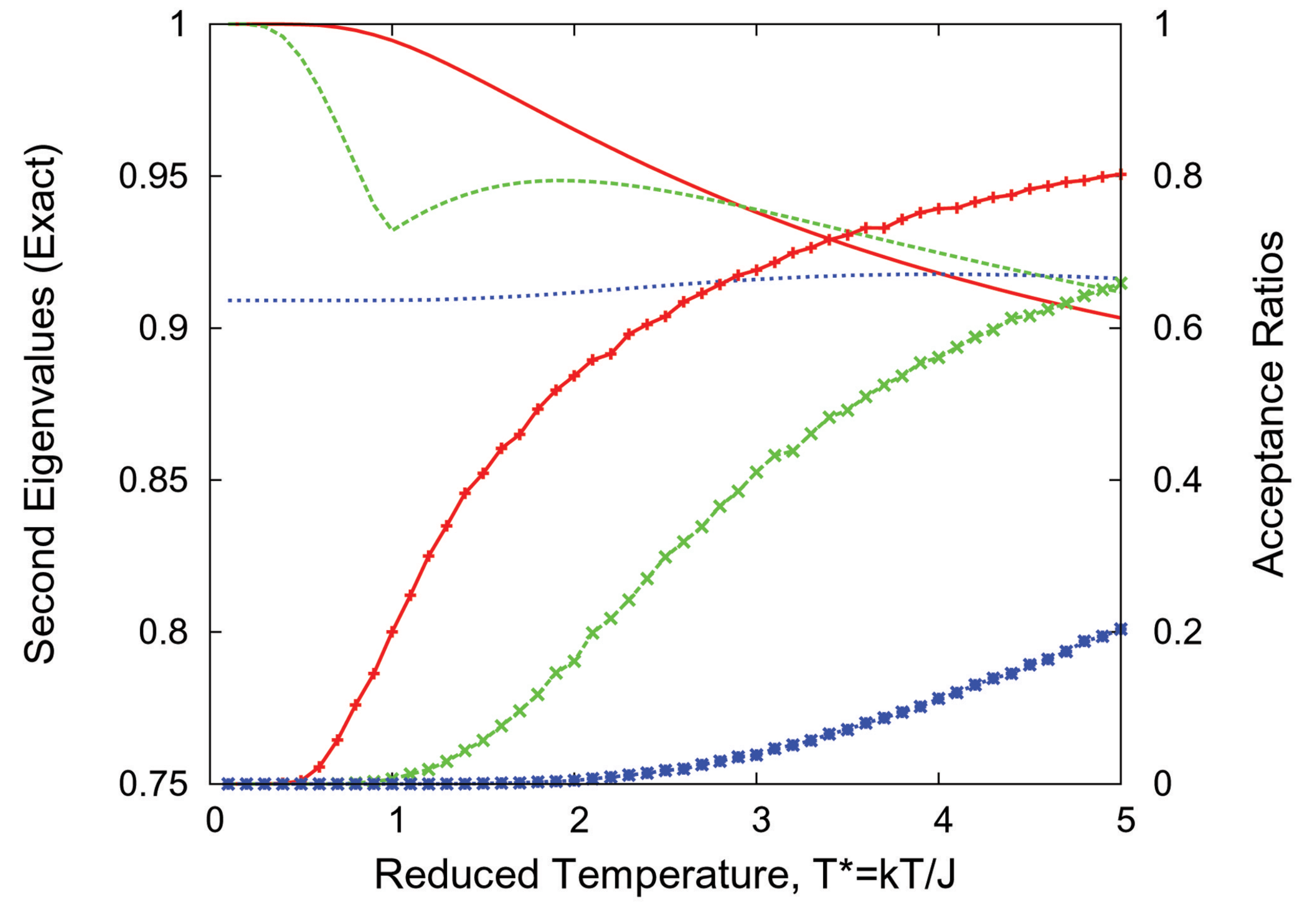}
\includegraphics[width=3in]{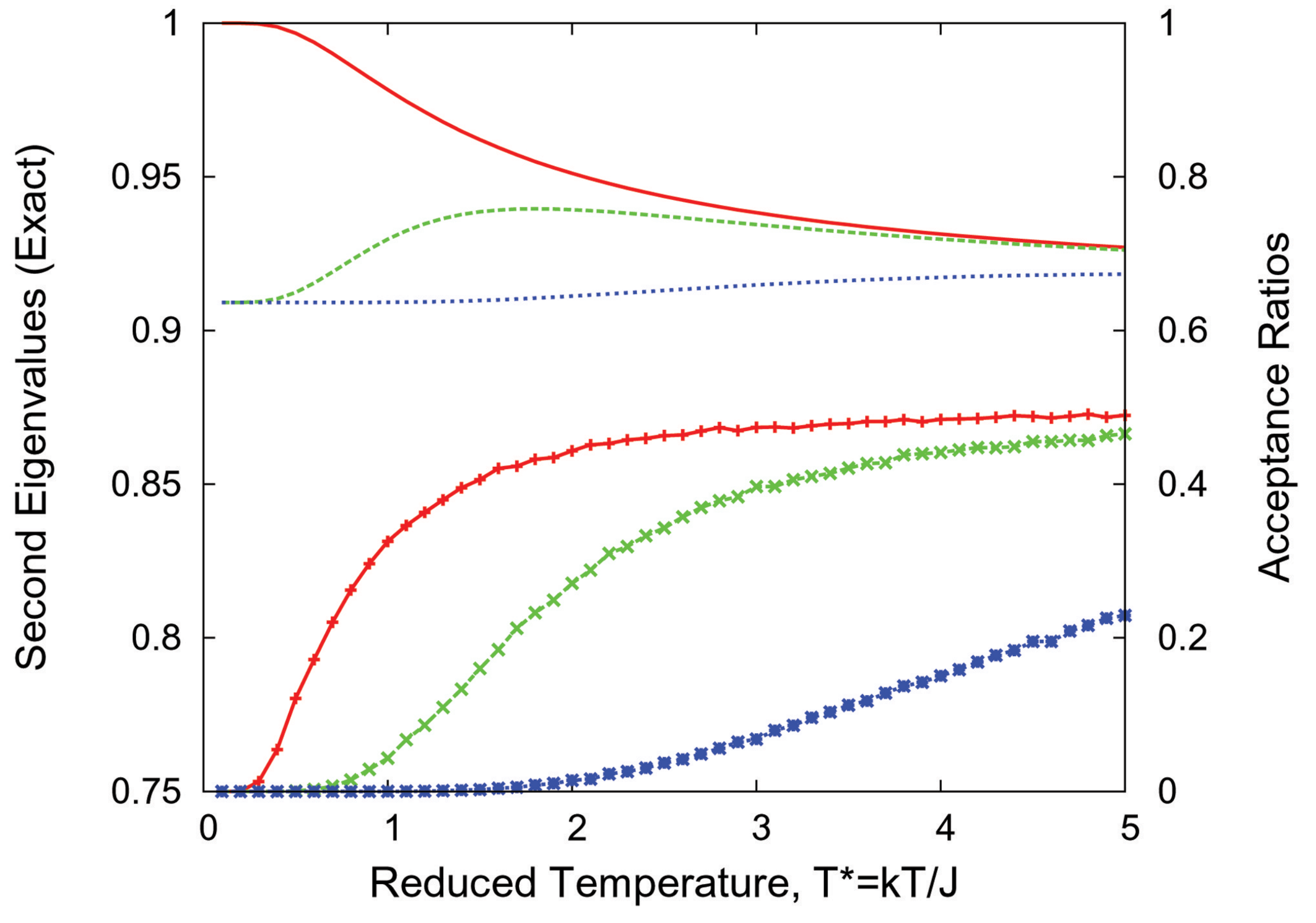}
\caption{Deterministic calculations of the second eigenvalues of the Metropolis (left) and heat bath (right) algorithms for the 10-site Ising lattice as a function of reduced temperature and magnetic field.  Red is for $H/J=0$; blue, for $H/J=1$; and green, for $H/J=4$.}
\label{fig4}
\end{center}
\end{figure}

In Fig.~\ref{fig2}a, we show the second eigenvalue for the Metropolis algorithm computed for short lattices deterministically and longer lattices stochastically, and in Fig.~\ref{fig2}b we show the same for the single-site heat-bath algorithm. As in Fig.~\ref{fig1}, the Metropolis's algorithm eigenvalue for reduced temperatures greater than 1 is always lower than that of the heat-bath algorithm for a given lattice size. For a given length and any algorithm, the magnitudes of the eigenvalues always decrease monotonically with increasing temperature. For reduced temperatures less than one, the eigenvalues closely approach unity.  The Metropolis algorithm approaches unity faster than the heat-bath algorithm.  While not shown, we remark that the second eigenvalue of all algorithms increase with lattice size, pushing the approach to unity to higher reduced temperatures. The difference in the temperature dependences of the eigenvalues and acceptance ratios for the Metropolis, singe-site heat bath, and two-site heat bath algorithms is shown in Fig.~\ref{fig3}. We note the high temperature crossover in the behaviors of the Metropolis and heat bath algorithms.

When the external field becomes non-zero, the high temperature lattice size trends are similar to the high temperature trends of the zero-field case. The low temperature behavior  however changes. In Fig.~\ref{fig4} we see that for both the Metropolis and heat bath algorithms the eigenvalue no longer approaches unity for sufficiently large fields at low temperatures. Further, the magnitude of the eigenvalue loses its monotonic decline as a function of temperature.

\begin{figure}[t]
\begin{center}
\includegraphics[width=3in]{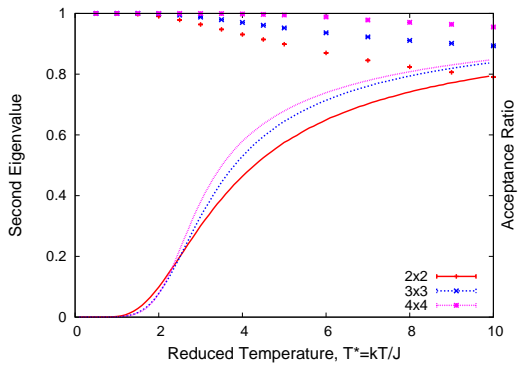}
\caption{The behavior of the second eigenvalues and acceptance ratio of the transition matrix for Metropolis algorithm as a function of the reduced temperature when applied to a $2 \times 2$, $3 \times 3$, and $4 \times 4$ Ising models in a zero magnetic field. The markers record the results for the eigenvalue, and the solid line, for the acceptance ratio.}
\label{fig6}
\end{center}
\end{figure}

In Fig.~\ref{fig6} we show results for a $2 \times 2$, $3 \times 3$, and $4 \times 4$ zero-field Ising model for the Metropolis algorithm. The markers are the eigenvalue results, and the solid line, those for the acceptance ratios. The behavior of the second eigenvalue here is quite consistent with the behavior observed in the one-dimesional case. As the reduced temperature is lowered, the second eigenvalue approaches unity. As the system size increases, the approach occurs at higher and higher reduced temperatures. In the thermodynamic limit, the two-dimensional Ising model has a well known critical temperature at $T_c^*=2.2692$. For the $4 \times 4$ lattice the second eigenvalue is effectively unity well above this temperature. The loss of efficiency in sampling for Ising models by the Metropolis (and heat-bath) algorithms as temperature is lowered is more a consequence of the increasing inefficiency of the algorithms than that of approaching a critical point. Results for the non-zero field case are qualitatively similar to those for the non-zero field one-dimensional Ising model.

In the Ising simulations, the accuracy of our results is principally controlled by the number of walkers used. Typical numbers ranged from 10,000 to 50,000, as the number of sites were varied from 9 to 25. The number of walkers was held constant over the range of temperatures simulated for a given lattice size and field. 

\begin{figure}[t]
\begin{center}
\includegraphics[width=3in]{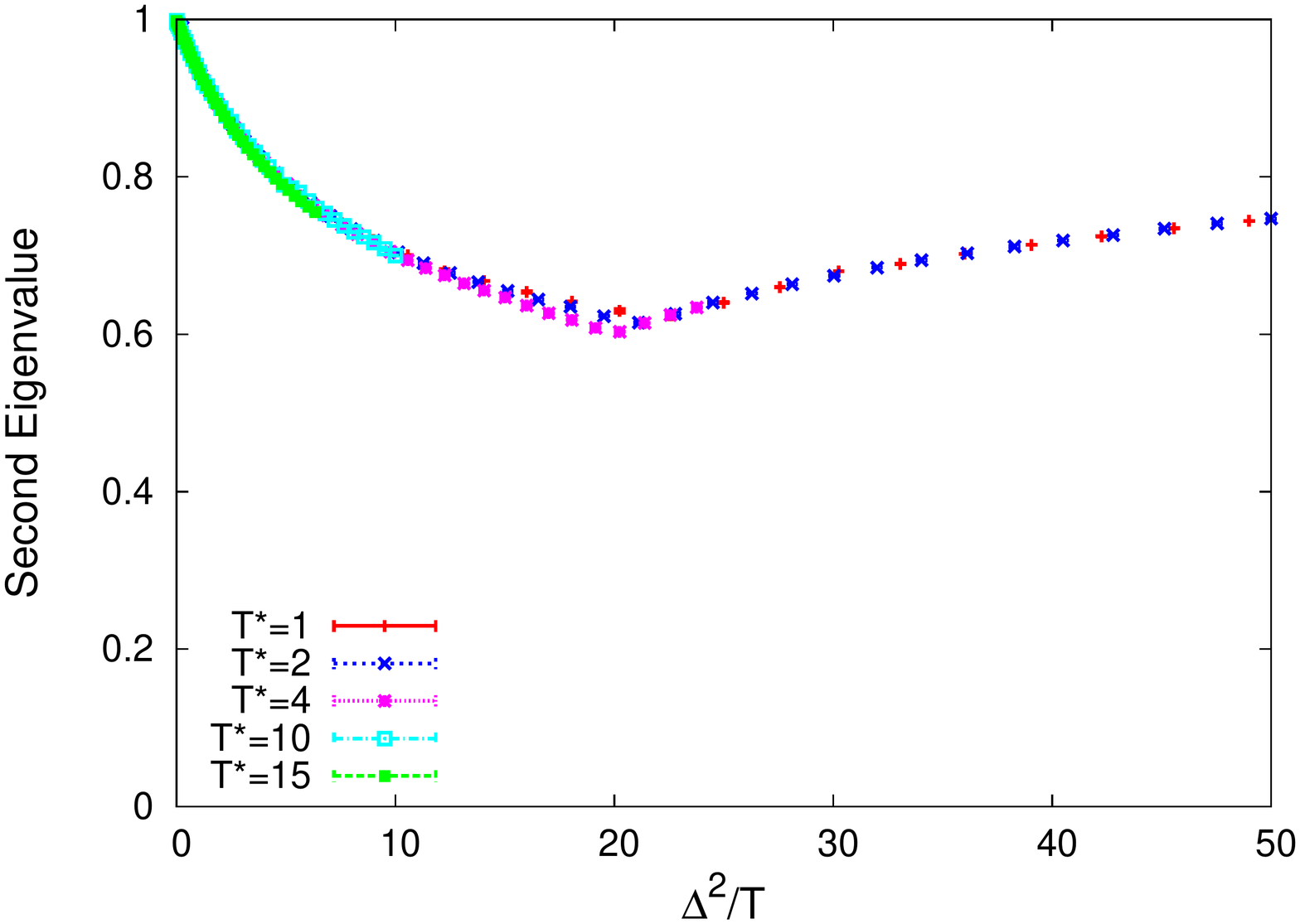}
\includegraphics[width=3in]{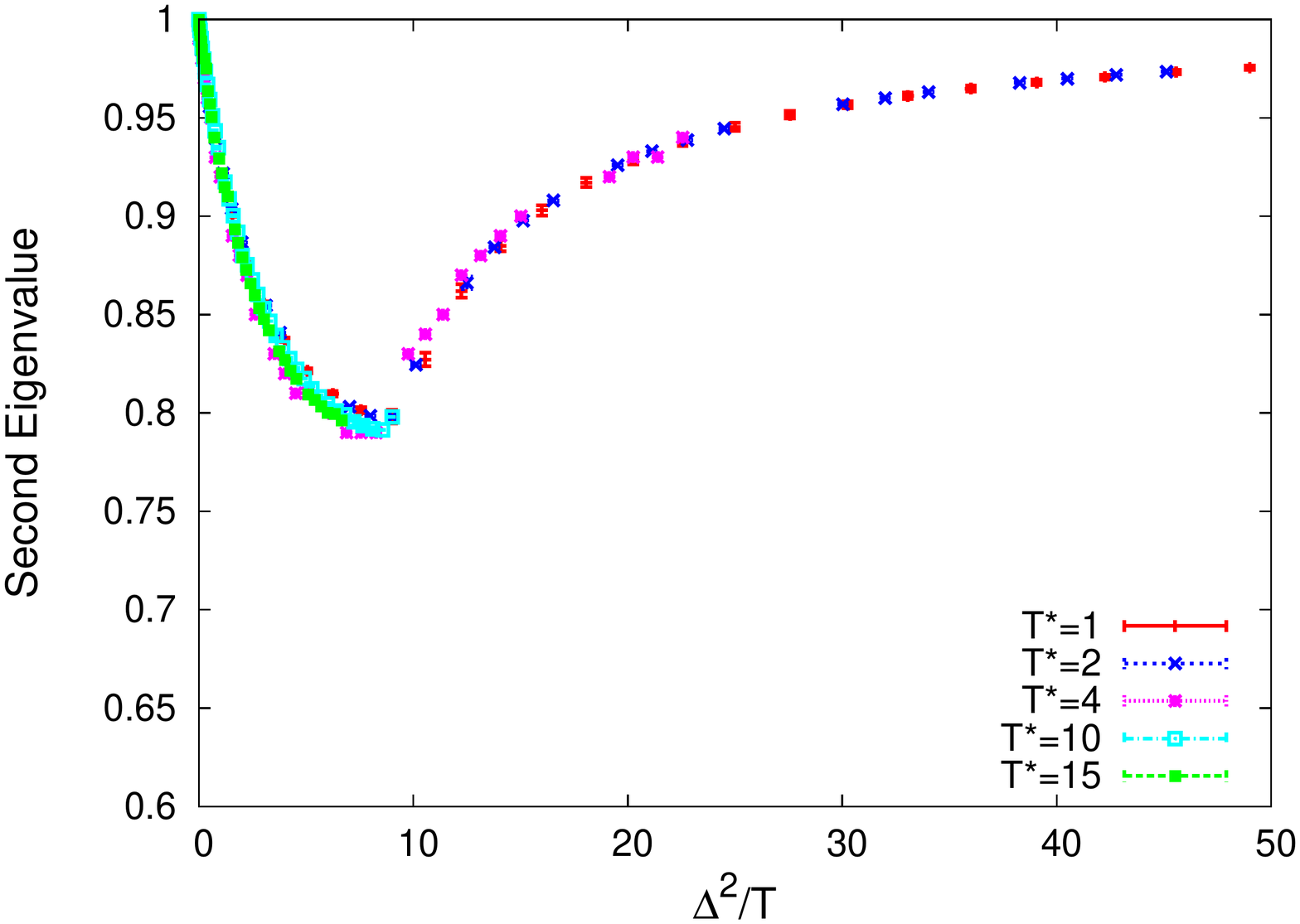}
\caption{The scaling of the second eigenvalue for the Metropolis transition matrix as a function of $\Delta^2/T$ for $N=1$ (left) and $3$ (right) non-interacting particles in the harmonic trap. $K=1$.}
\label{fig7}
\end{center}
\end{figure}

\subsection{Harmonic Trap}

We started by considering one particle in the harmonic trap to benchmark our basic Monte Carlo procedures against the deterministic single harmonic oscillator results of Doll \emph{et al.} \cite{doll09}, an equivalent problem approached with the use of standard eigenvalue software. In terms of (\ref{eq:harmonic_trap}), we took $N=1$ and $C=0$. We discretized the one-dimensional space, following the prescription of Doll {\em et al.}, selected a Metropolis box size and potential-dependent cutoffs, and computed the transition matrix $P_{X',X}$. \emph{Both the distance $\delta$  proposed for the walker to move and the Metropolis box size $\Delta$, were discrete multiples of an underlying cell width, which was 0.05. }  In units of the cell width our cut-off distance $\Gamma=400$, which was more than sufficient for most temperatures. We again diagonalized the transition matrix using standard eigensystem software and found excellent agreement between our Monte Carlo determination, performed with the same Monte Carlo techniques used for the Ising model, and our deterministic results.

To test our continuum Monte Carlo method, we exploited the observation of  Doll \emph{et al.}, proven in our Appendix, that for a power-law potential $x^n$, all the eigenvalues of the Metropolis $P_{X',X}$ are a function solely of $\Delta^n/T$. As illustrated in Fig.~\ref{fig7}, we achieve excellent scaling over a wide range of reduced temperatures with $N=1$ ( the Doll case) and $N=3$ with $C=0$ and $K=1$.  We note a characteristic feature of the second eigenvalue of the single oscillator is a minimum value. In going from the discretized case to the continuum, this value decreases. As we can see from Fig.~\ref{fig7} adding more non-interacting particles to the trap increases the continuum value.

\begin{figure}[p]
\begin{center}
\includegraphics[width=2.9in]{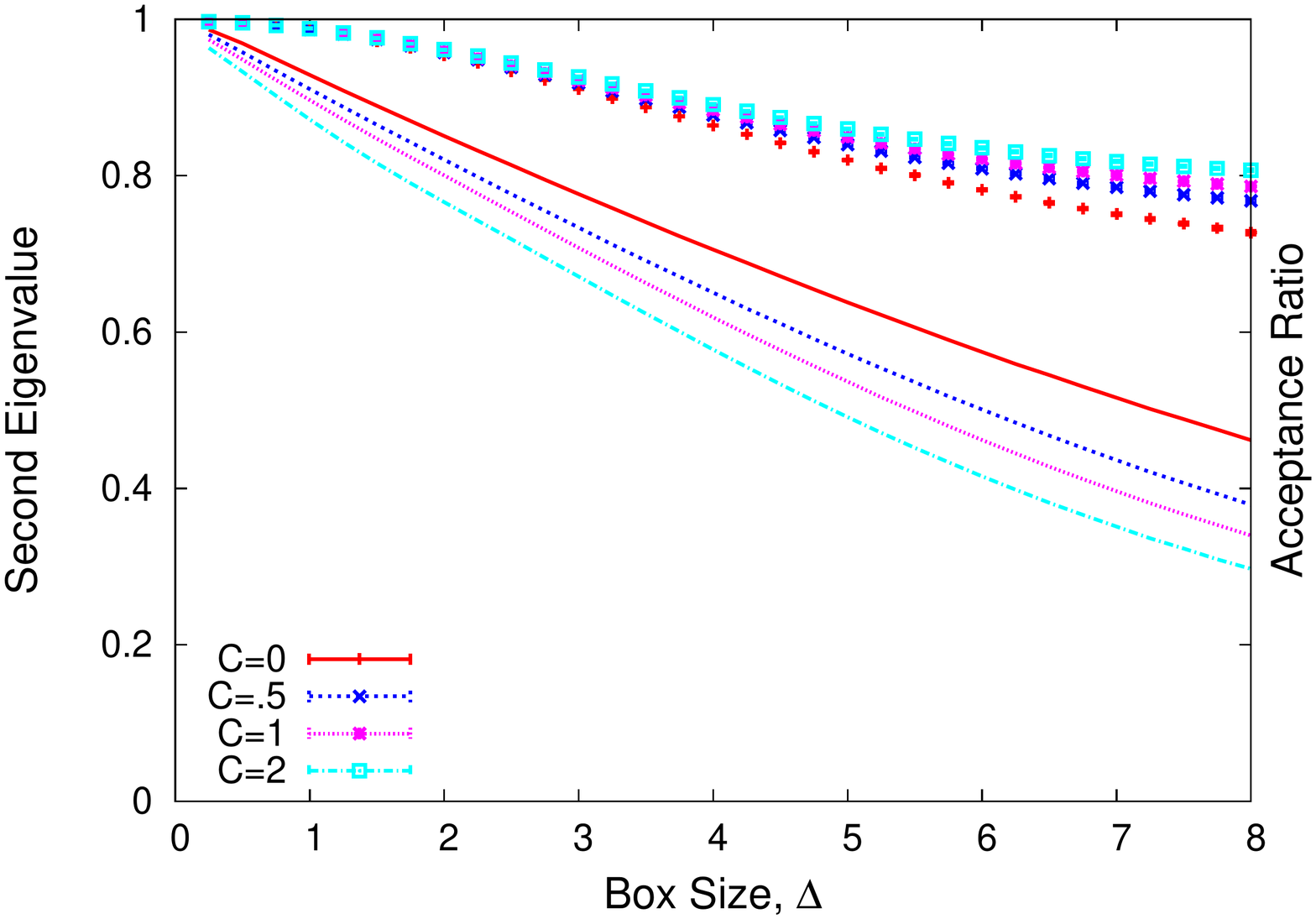}
\includegraphics[width=2.9in]{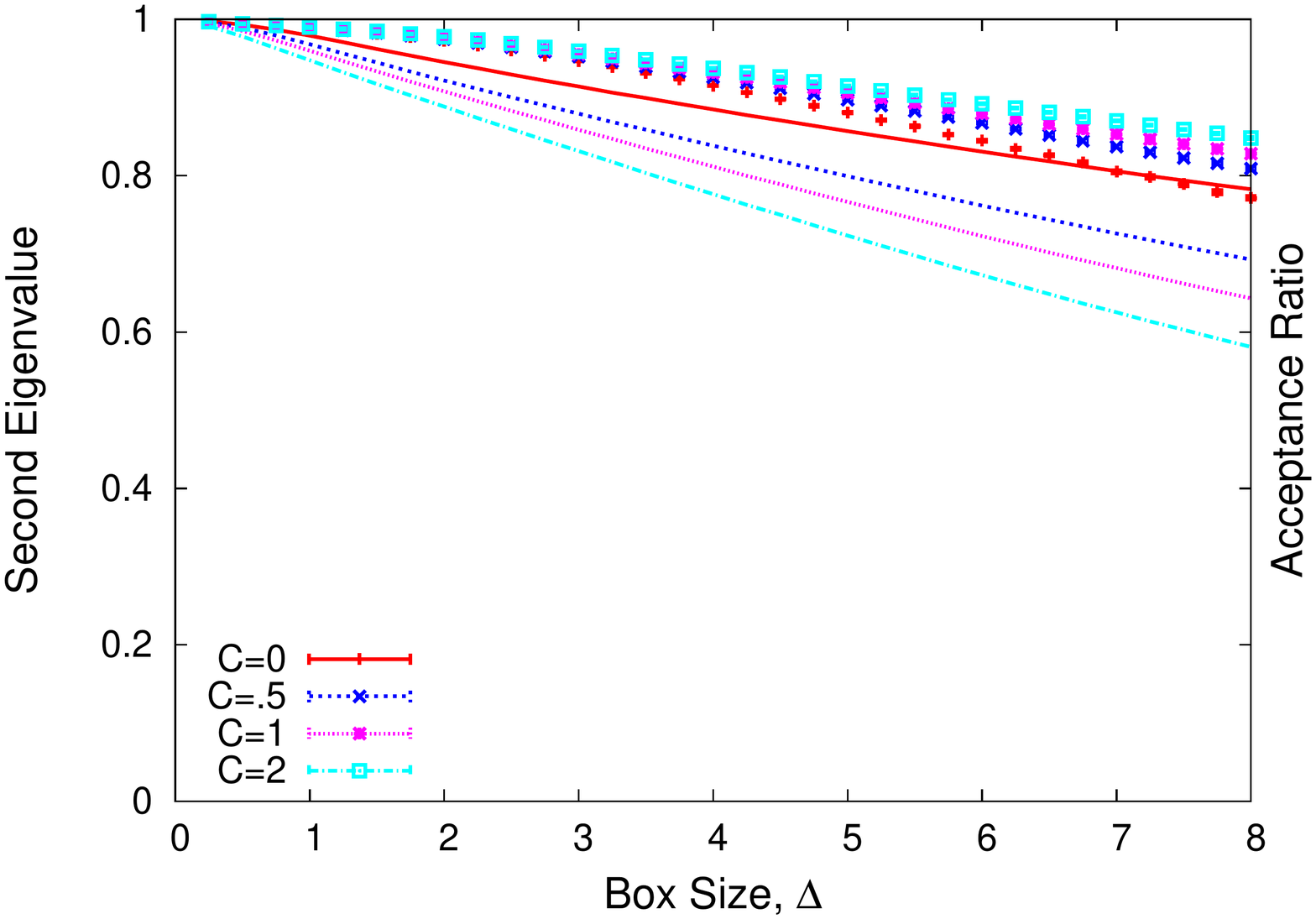}
\includegraphics[width=2.9in]{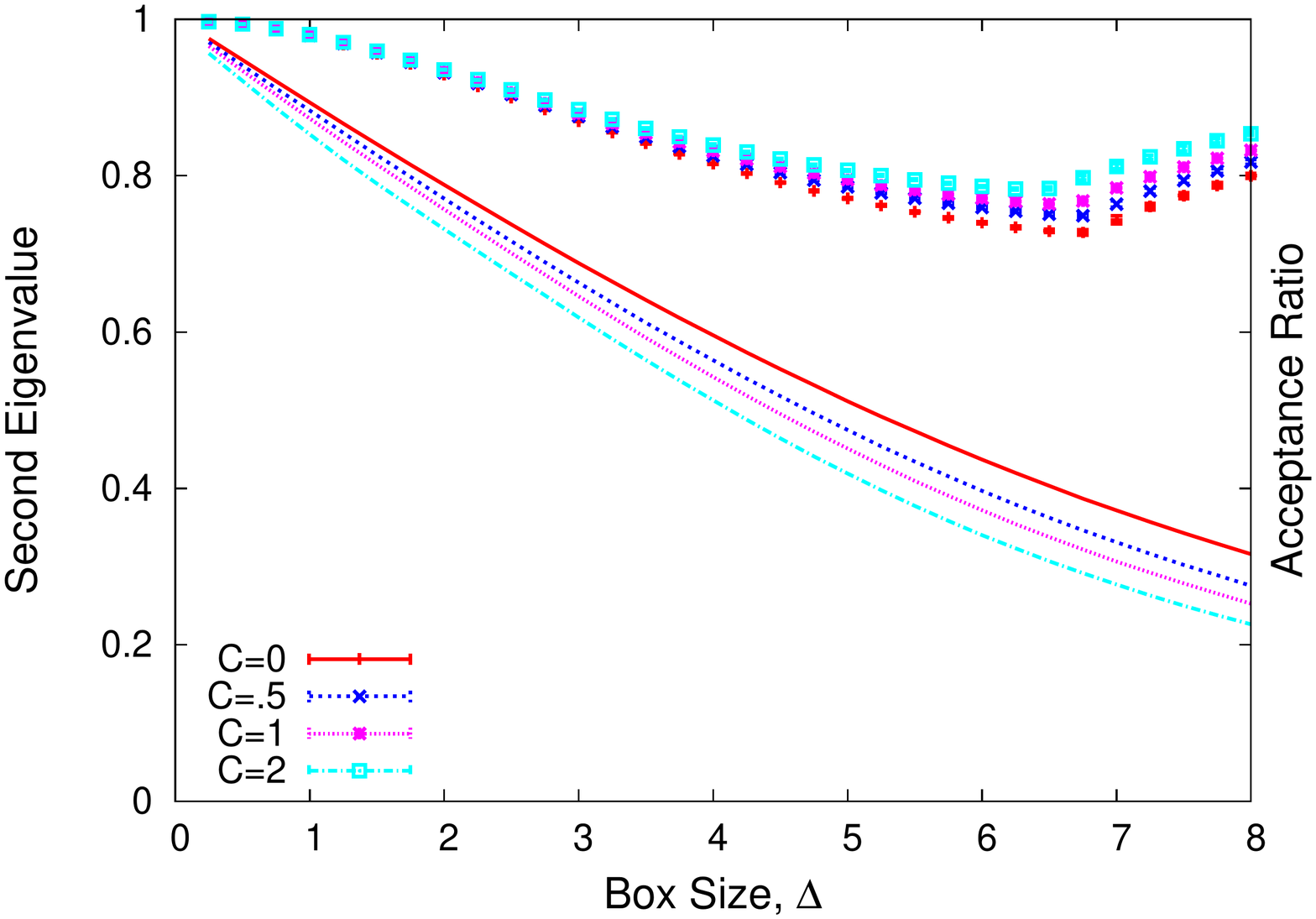}
\includegraphics[width=2.9in]{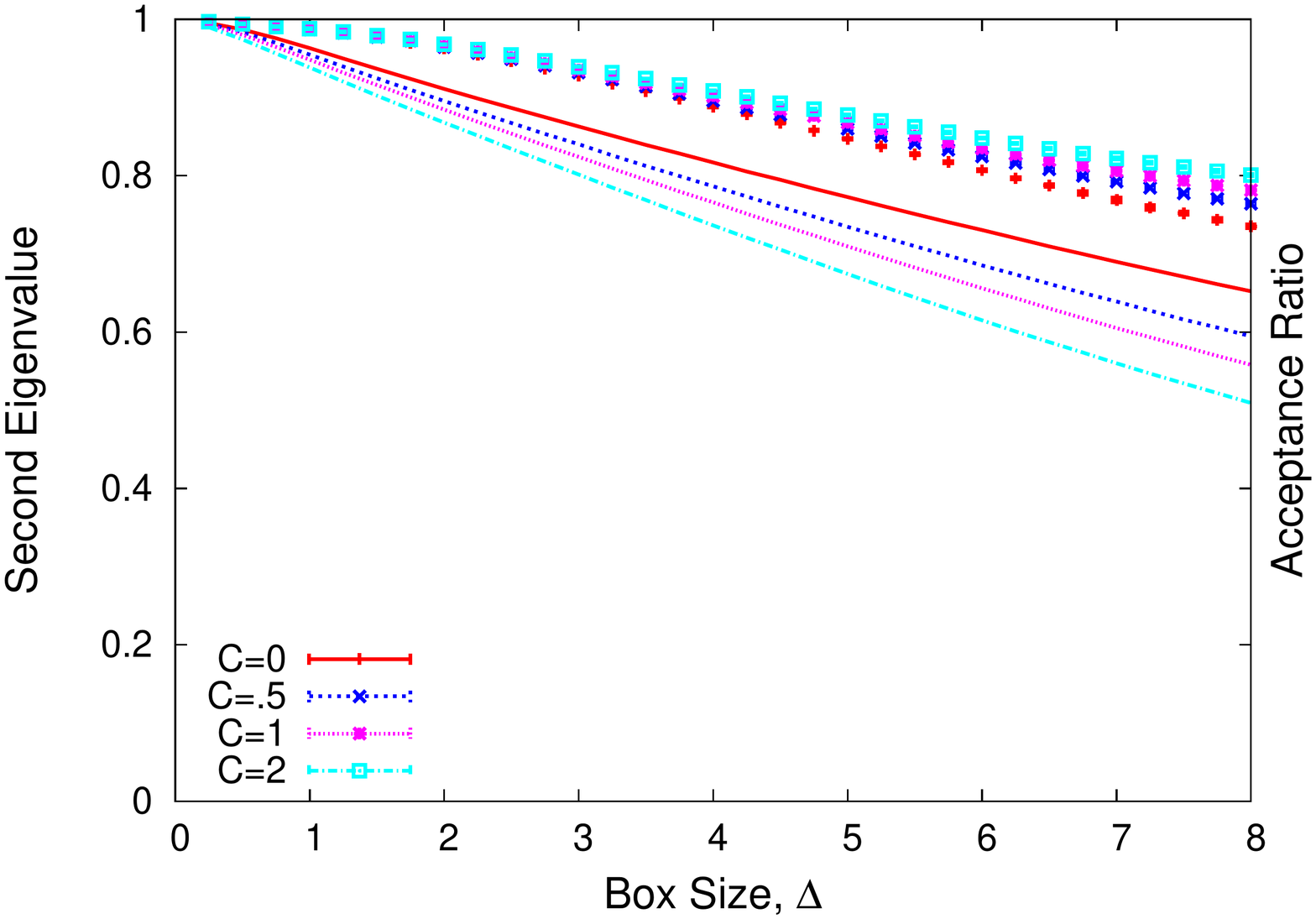}
\includegraphics[width=2.9in]{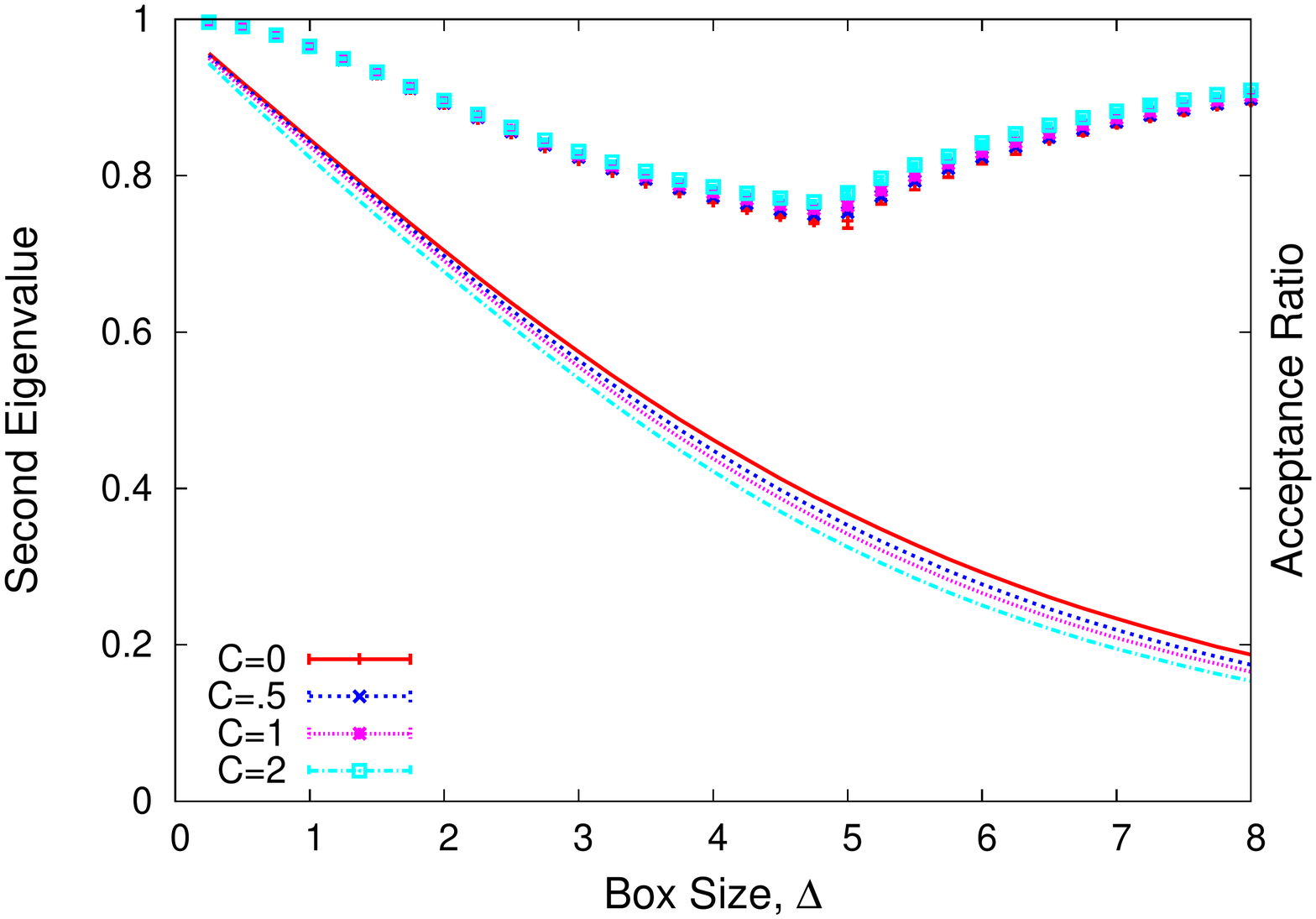}
\includegraphics[width=2.9in]{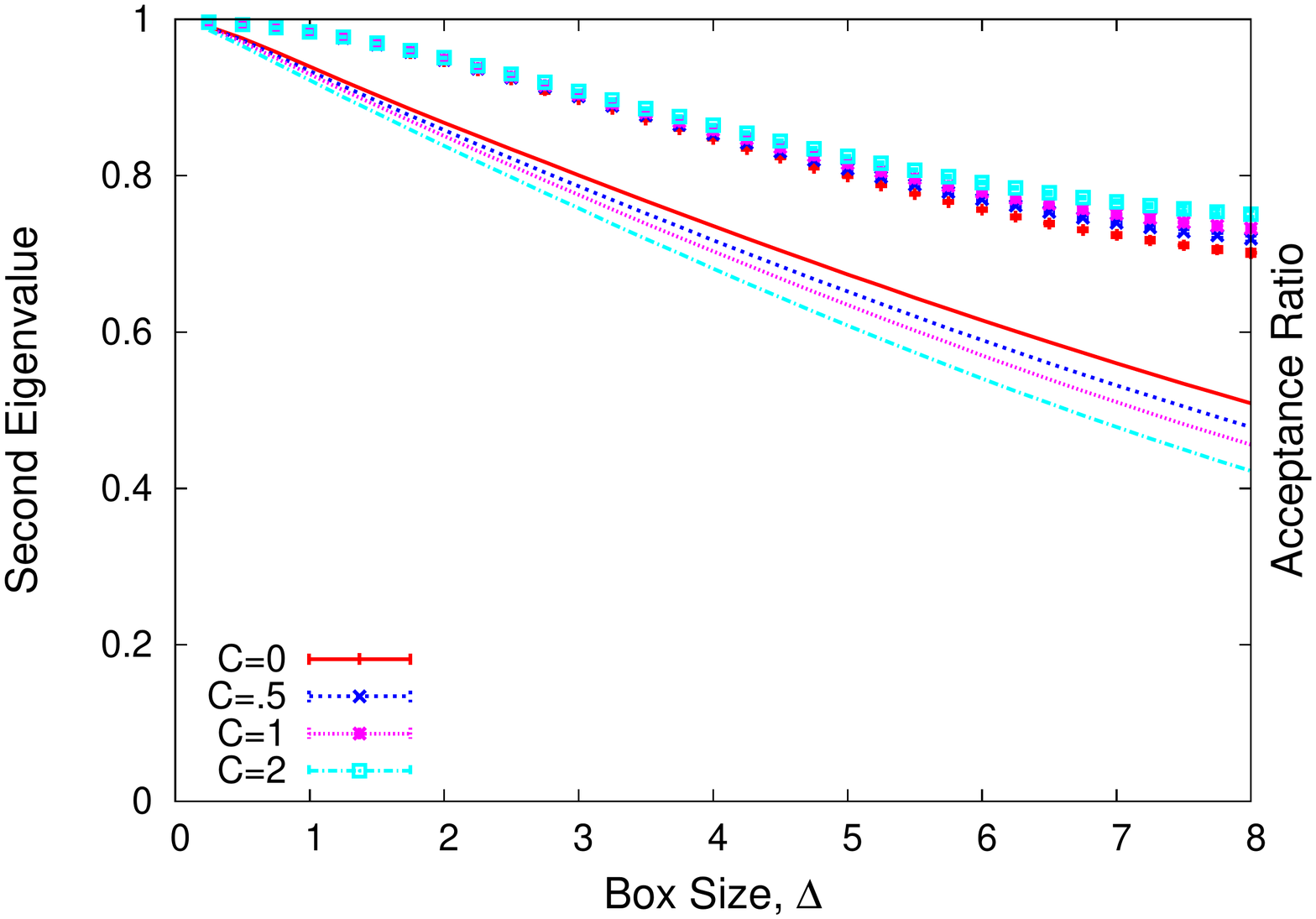}
\caption{The second eigenvalues and acceptance ratios for the Metropolis algorithm  as a function of Metropolis box size $\Delta$ for two particles in the harmoninc trap for various values of the coupling constant $C$. Down the left column $T^*=2$; down the right, $T^*=10$. Across the top row $K=0.5$; the middle row, $K=1$; and the bottom row, $K=2$.}
\label{fig8}
\end{center}
\end{figure}

\begin{figure}[t]
\begin{center}
\includegraphics[width=3in]{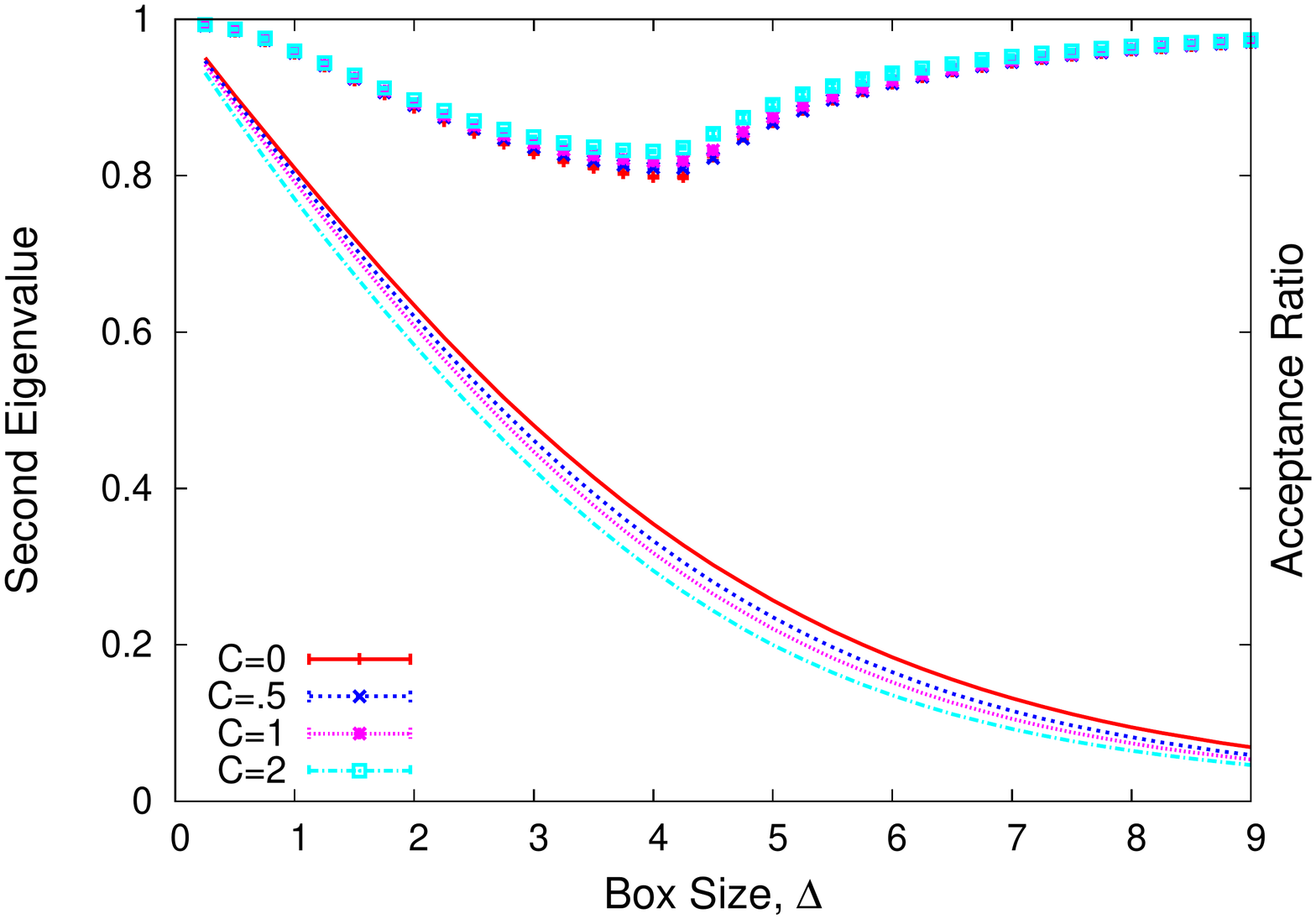}
\includegraphics[width=3in]{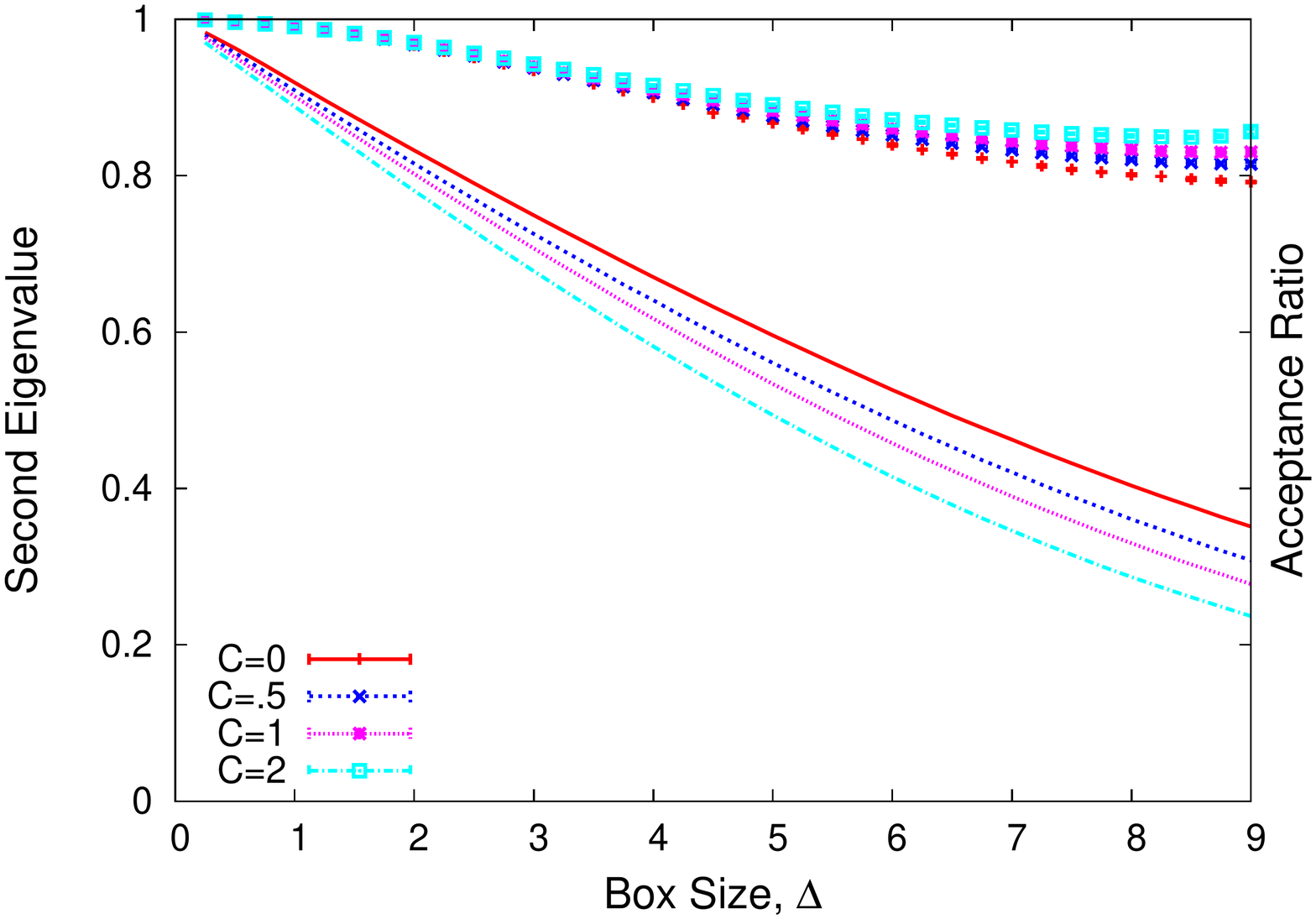}
\caption{The second eigenvalue for the Metropolis transition matrix for 3 particles in the harmonic trap for various values of their coupling constant $C$ as a function of Metropolis box size. $K=2$   On the left, $T^*=2$, while on the right $T^*=10$.}
\label{fig9}
\end{center}
\end{figure}

In Fig.~\ref{fig8} we present a coarse summary of the behavior of the second eigenvalue and acceptance ratio as a function of box size for a two-particle trap at a low and high temperature for several values of the trap curvature $K$ and various values of the coupling $C$ between particles. In all cases, the acceptance ratio uniformly decreases as  a function of box size. Its sensitivity to the coupling $C$ depends on $K$. It increases as $C$ increases, less so for the larger values of $K$. At $T^*$, the sensitivity of the second eigenvalue on $C$ anti-correlates with the sensitivity of the acceptance ratio, showing more sensitivity at the smaller values of $K$ and decreasing with increasing $C$. At $T^*=2$, many of the eigenvalues trends are the same as those for $T^*$. The key difference is the absence of a minimum value of the eigenvalue for $K=0.5$ but its present at the other two $K$ values and its location shifting toward smaller box sizes as $K$ increases. A minimum occurs for $K=0.5$ but is located at a box size larger than we simulated. 

Our final results are shown in Fig.~\ref{fig9}. Here, we have $N=3$ and a single value of the curvature, $K=2$. The results show the same trends as the $N=2$ and $K=2$ case in Fig.~\ref{fig8}. The minimum in the eigenvalue however shifted towards smaller box size.

Figures \ref{fig7}-\ref{fig9} suggest that at high temperature, the box size, which we recall is measured in units of the cell width,  should be as large as possible. This seems reasonable \cite{note3}. We note the acceptance ratio will then be below the $40\%$. At low temperature, there is typically  an optimal box size. For the cases presented, the acceptance ratio is around $40\%$.

Besides the number of walkers, the accuracy of the results also depends on the cell width. The value of 0.05 was found by experimentation to be convenient both with respect to accuracy and efficiency. Typical numbers of walkers ranged from 50,000 to 200,000 as the number of particles was increased from one to three. Fewer walkers could be used for higher temperatures than for lower ones for the same system.

\section{Concluding Remarks}

We proposed and benchmarked a numerical method for computing the subdominant eigenvalue $\lambda_2$ of a matrix or continuous operator. Based on the work of \cite{booth09} and \cite{yamamoto09}, this method can be implemented deterministically and stochastically, computes just this one eigenvalue, and requires knowing the dominant eigenpair. For Markov chain transition matrices, this pair is known analytically so we used the method to compute the $\lambda_2$ of various transition matrices.  For such matrices $0<|\lambda_2|<1$, with small $|\lambda_2|$ implying large Monte Carlo efficiency.  Specifically, we computed the $\lambda_2$ of the transition matrices for several one and two dimensional Ising models, which have a discrete phase space, and compared the relative efficiencies of the Metropolis and heat-bath algorithms as a function of temperature and applied magnetic field. Based on the $\lambda_2$ criterion, we found that small lattices appear to give an adequate picture of comparative efficiency and that the heat-bath algorithm is more efficient than the Metropolis algorithm only at low temperatures where both algorithms are inefficient. We also computed the $\lambda_2$ of the transition matrix of a model of an interacting gas trapped by a harmonic potential, which has a mutidimensional continuous phase space, and studied the efficiency of the Metropolis algorithm as a function of temperature and the maximum allowable step size $\Delta$.  We found that the traditional rule-of-thumb of adjusting $\Delta$ so the Metropolis acceptance rate is around 50\% range is often sub-optimal. In general, as a function of temperature or $\Delta$,  $\lambda_2$ for this model displays trends defining optimal efficiency that the acceptance ratio does not. The cases studied also suggest that Monte Carlo simulations for a continuum model in the continuum are likely more efficienct than those for a discretized version of the model. 

Many of our results and conclusions, of course, could be limited to these specific cases; however, we suggest establishing their degree generality would be advantageous. For example, we quantified situations where focusing on acceptance ratios is ill advised. From our experience, this result is not very surprising. Clearly, computing $\lambda_2$ is more justified and gave more specific information about which of several algorithmic approaches or adjustments is likely preferable. More interestingly, if we can always capture efficiency trends for small systems sizes, this is a significant simplification.

Our Monte Carlo techniques were simple. In treating more complicated problems, we might need some of the methods used in \cite{booth09,booth09b} plus others. Our techniques and approach were also quite different than those by \cite{nightingale96a,nightingale96b,nightingale00}.  For example, our starting functions were considerably simpler to construct, and we did not need to use these functions as an importance function guiding our random walkers. Additionally, we used exact estimators for the eigenvalue estimates instead of variational ones. On other hand, we must be concerned with the cancellation of positively and negatively signed walkers. Our procedures for doing so however where quite simple and effective, and our success means we solved a type of sign problem. We do not expect our eigenvalues to be as precise as is possible with this other method.  In most cases, our eigenvalues estimate errors  were smaller than marker sizes. For present high precision in these estimates is  unnecessary. More advantageous starting functions could easily be used if needed.

The most substantive issue for future work is more fully understanding  the utility of $\lambda_2$ as a metric for comparative algorithmic efficiency. Focusing on this quantity has the advantage of it being well defined and backed by some rigorous results.  From a practical point of view, the most efficient algorithm is the one that produces a measurement of a required accuracy with the less amount of computer time. Using $\lambda_2$ does not address the time is takes to perform a Monte Carlo step; it only says something about the number of steps needed. Experience has shown that the fast or slow relaxation of an algorithm is generally accompanied by fast of slow generation of statistically independent measurements. Likely a good $\lambda_2$ is a necessary but not sufficient rule-of-thumb. We note that Peskun's result (\ref{eq:peskun}) insures the variance of any measurable of algorithm 1 will be less than of equal to that computed from algorithm 2. The connection of $\lambda_2$ to his result about measurables is indirect.

\appendix*
\section{Scaling}

The basic equation for a Markov chain is
\[
p(x) = \int {dy\,P(x,y)p(y)} 
\]
Here, $p$ is the stationary probability density of the chain and $P$ is the transition probability density defining the chain. They satisfy
\begin{eqnarray*}
                               \int dx\,p\left( x \right) &= 1  \\ 
	 \int dx\,P\left( {x, y} \right) &= 1   
\end{eqnarray*}
The Metropolis algorithm makes a specific choice for the transition probability density
\[
P\left( {x, y} \right) = T\left( {x, y} \right)\min \left\{ {1,{{\exp \left[ { - \beta V\left( y \right)} \right]} \mathord{\left/
 {\vphantom {{\exp \left[ { - \beta V\left( y \right)} \right]} {\exp \left[ { - \beta V\left( x \right)} \right]}}} \right.
 \kern-\nulldelimiterspace} {\exp \left[ { - \beta V\left( x \right)} \right]}}} \right\}
\]
where
$
\int {dx\,T\left( {x,y} \right) = 1} .
$
For the proposal density  $T( x , y )$ the Metropolis choice has it non-zero only on the interval 
$
\left[ {y - {\textstyle{\Delta  \over 2}},y + {\textstyle{\Delta  \over 2}}} \right]
$
over which its magnitude is $1/\Delta$.  We will show that for potentials of the form $x^n$ the product of this choice of the proposal probability density and the standard Metropolis acceptance density
\[
A(x,y)=\min \left\{ {1,{{\exp \left[ { - \beta V\left( y \right)} \right] } \mathord{\left/
 {\vphantom {{\exp \left[ { - \beta V\left( y \right)} \right]} {\exp \left[ { - \beta V\left( x \right]} \right)}}} \right.
 \kern-\nulldelimiterspace} {\exp \left[{ - \beta V\left( x \right)} \right]}}} \right\}
\]
lead to a scaling of $P(x,y)$ that implies all it eigenvalues scale as a function of $\Delta^n/T$

We will measure all lengths in units of $\Delta$ and start by writing the acceptance function as 
$
A(x,y) = A(x,y;T)
$
to make the $T$ dependence explicit. Next, for potentials of the type $x^n$
\begin{eqnarray*}
 A(x,y;T) &=& \min \left\{ {1,{{\exp \left( { - \frac{{y^n }}{T}} \right)} \mathord{\left/
 {\vphantom {{\exp \left( { - \frac{{y^n }}{T}} \right)} {\exp \left[ { - \frac{{x^n }}{T}} \right]}}} \right.
 \kern-\nulldelimiterspace} {\exp \left( { - \frac{{x^n }}{T}} \right)}}} \right\} \\ 
&=& \min \left\{ {1,{{\exp \left[ { - \frac{{\left( {\frac{y}{\Delta }} \right)^n }}{{\left( {\frac{T}{{\Delta ^n }}} \right)}}} \right]} \mathord{\left/
 {\vphantom {{\exp \left[{ - \frac{{\left( {\frac{y}{\Delta }} \right)^n }}{{\left( {\frac{T}{{\Delta ^n }}} \right)}}} \right]} {\exp \left[ { - \frac{{\left( {\frac{x}{\Delta }} \right)^n }}{{\left( {\frac{T}{{\Delta ^n }}} \right)}}} \right]}}} \right.
 \kern-\nulldelimiterspace} {\exp \left[ { - \frac{{\left( {\frac{x}{\Delta }} \right)^n }}{{\left( {\frac{T}{{\Delta ^n }}} \right)}}} \right]}}} \right\} \\
&=& A\left( {\frac{x}{\Delta },\frac{y}{\Delta };\frac{T}{{\Delta ^n }}} \right) 
\end{eqnarray*}
Thus, from the point of view of the acceptance, going from $y$ to $x$ at temperature $T$ is the same as going from $x/\Delta$ to $ y/\Delta$ at $T/\Delta^n$.

We now write 
$
T\left( {x, y} \right) = T\left( {x, y;\Delta } \right)
$
which distributes $x$ uniformly over an interval centered at $y$ and of width $\Delta$, that is,
\[
x \in \left\{ {y - \frac{\Delta }{2} \le y \le y + \frac{\Delta }{2}} \right\}
\]
Over this interval its amplitude is $1/\Delta$.  Next, we consider
\[
\frac{x}{\Delta } \in \left\{ {\frac{y}{\Delta } - \frac{1}{2} \le \frac{y}{\Delta } \le \frac{y}{\Delta } + \frac{1}{2}} \right\}
\]
A function that distributes $x/\Delta$ uniformly over this unit interval and has a unit amplitude will make an acceptable proposal probability. Such a function is
$
\Delta \,\,T\left( {\frac{x}{\Delta }, {\frac{y}{\Delta }};1} \right).
$
Thus for the scaled system we have
\begin{eqnarray*}
 P\left( {\frac{x}{\Delta }, {\frac{y}{\Delta };1,\frac{T}{{\Delta ^n }}}} \right) 
&=& \Delta \,\,T\left( {\frac{x}{\Delta }, {\frac{y}{\Delta };1}} \right)A\left( {\frac{x}{\Delta },\frac{y}{\Delta };\frac{T}{{\Delta ^n }}} \right) \\ 
&=& T\left( {x, {y;\Delta }} \right)A\left( {x,y;T} \right) \\ 
&=& P\left( {x, {y;\Delta ,T}} \right)
 \end{eqnarray*}
which establishes the scaling of all the eigenvalues of $ P\left( {x,{y;\Delta ,T} } \right)$

%\begin{acknowledgments}
%\end{acknowledgments}



\begin{thebibliography}{}
\bibitem{note1} Here, our definition of the second eigenvalue as $\lambda_2$ differs for the choice of \cite{doll09} which is $1-\lambda_2$.
\bibitem{note2} As a guide to the choice of regions, we comment left and right eigenstate corresponding to different eigenvalue are orthogonal. The subdominant state we seek is thus orthogonal to the dominant right-hand state which is the Boltzmann distribution. Likely good choice of regions will reflect regions with high Boltzmann weight.
\bibitem{note3} Because the eigenvalues depend on box size, it is very possible that there are minima in these graphs, only at very large box sizes. So, it is likely that one should not increase the box size endlessly, just increasing it at high temperatures more than at low temperatures.
\end{thebibliography}
\end{document}